\documentclass[%
 reprint,
 amsmath,amssymb,
 aps,
]{revtex4-2}

\usepackage{graphicx}
\usepackage{dcolumn}
\usepackage{bm}
\usepackage[table]{xcolor}
\usepackage{braket}

\begin{document}

\preprint{APS/123-QED}

\title{Hybrid Fiber-Free-Space Entanglement Distribution Using Off-the-Shelf Quantum Devices}

\author{Gustavo C. Amaral}
 \altaffiliation[Corresponding Author: ]{gustavo.castrodoamaral@tno.nl}
\author{Nienke M. ten Haaf}
\author{Breno Perlingeiro}
\author{David L. Bakker}
\author{Mark G. M. Boekel}
\author{Tim E. van Duivenbode}
\author{Karthik Selvan}
 \altaffiliation[Currently at ]{Netherlands Aerospace Centre}
\author{Nicolas Oidtmann}
 \altaffiliation[Currently at ]{Da Vinci Satellite Group}
\author{Rafael Ochsendorf}
\author{Rick N. M. Wasserman}
\affiliation{%
 The Netherlands Organization for Applied Scientific Research, Delft, The Netherlands
}%

\author{Mael Flament}
\author{Felipe Giraldo}
\author{Shane Andrewski}
\author{Mehdi Namazi}
\affiliation{
 Qunnect NL B.V., Delft, The Netherlands
}%

\author{Federica Facchin}
\author{Mario Castañeda}
\affiliation{%
 SingleQuantum B.V., Delft, The Netherlands
}%

\author{Fokko de Vries}
\author{Sayali Shevate}
\author{Shaurya Bhave}
\affiliation{%
 Qblox B.V., Delft, The Netherlands
}%

\author{Marco Gorter}
\author{Nico Coesel}
\affiliation{%
 OPNT B.V., Delft, The Netherlands
}%

\author{David Mytling}
\author{Mike Mabry}
\author{Carlo Page}
\author{Alexandra Pinto}
\affiliation{%
 Xairos B.V., Delft, The Netherlands
}%

\author{Joanneke Jansen}
\author{Rahul Vyas}
\author{Marc X. Makkes}
\affiliation{%
 Fortaegis Technologies B.V., Amsterdam, The Netherlands
}
\date{\today}

\begin{abstract}
Entanglement serves as a fundamental resource for quantum technologies, enabling communication and computation tasks that surpass classical limits. Its distribution across networks is essential for interconnecting quantum processors, enabling distributed quantum computing to address complex challenges in areas such as drug discovery, material science, and optimization. In this work, we report the successful distribution of polarization-entangled photon pairs across a campus-scale, three-node quantum network comprising both fiber and free-space optical links. The entire system was built using commercially available components provided by partners within the Netherlands Quantum Ecosystem. This result represents advancements in the technological maturity of quantum communication systems and demonstrates a pathway towards the practical deployment of early-stage quantum networks both on Earth and in space.
\end{abstract}

\maketitle

\section{Introduction}
The first quantum revolution brought technologies such as transistors and lasers by enabling control over quantum states in solid-state systems. The second quantum revolution harnesses uniquely quantum resources, especially entanglement, to surpass classical limits in computing, secure communication, sensing, and time transfer. Quantum computers, in particular, have the potential to solve problems considered intractable for classical systems, with the capacity to deliver major societal benefits and breakthroughs across multiple fields of science and technology. While today’s quantum computers remain in the Noisy Intermediate-Scale Quantum (NISQ) era \cite{preskill2018quantum}, ongoing advances in coherence, system integration, and error correction are steadily extending their capabilities. At present, their performance is primarily constrained by the limited scale of quantum processors. Overcoming this scaling challenge requires both improvements in the underlying physical platforms and the development of distributed quantum computing architectures, which are essential to bypass size limitations and connect heterogeneous systems.

\begin{figure*}[t!]
\centering
\includegraphics[clip, trim=0cm 3cm 1.45cm 3.3cm, width=\linewidth]{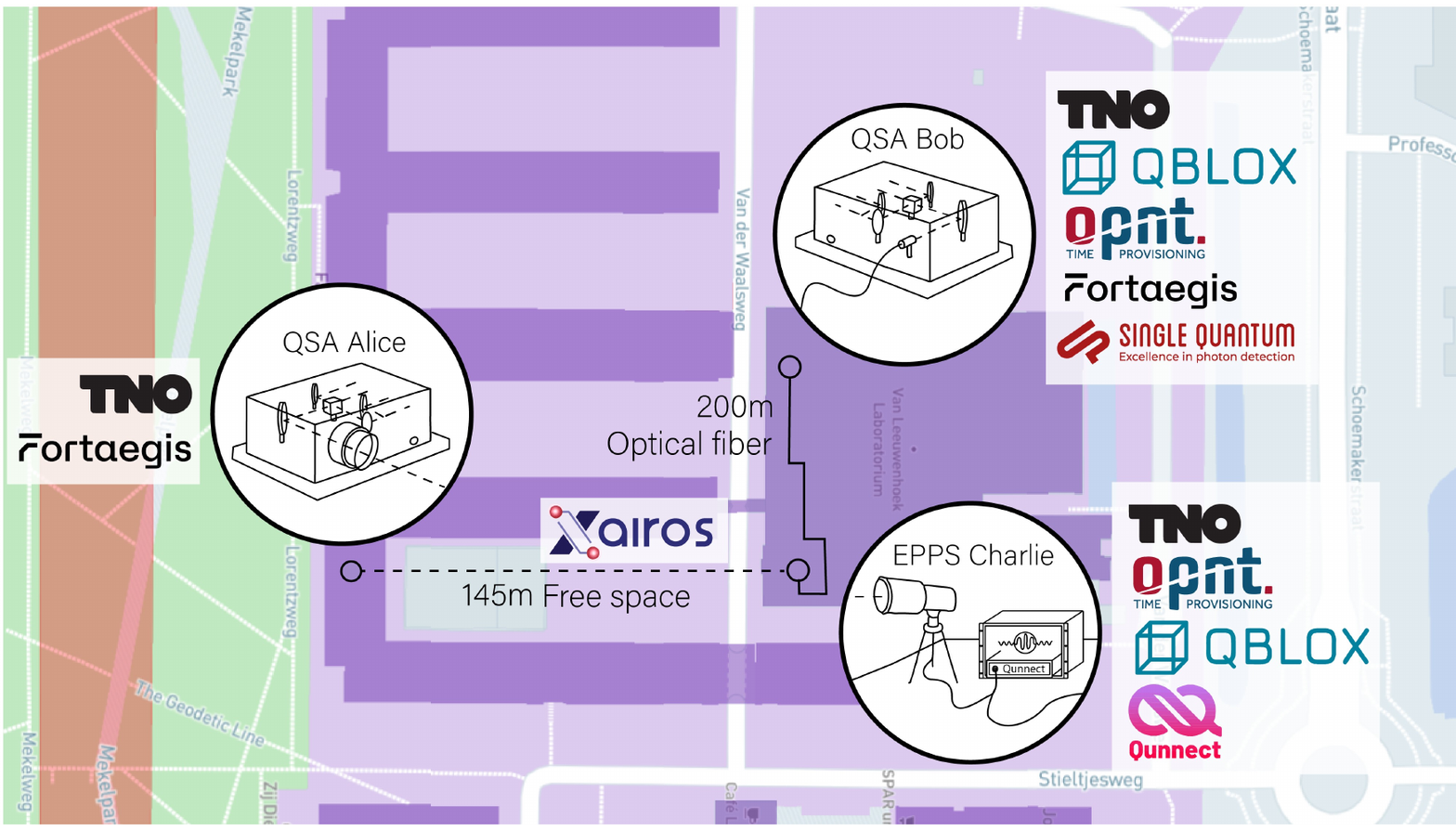}
\caption{Map of the Delft University of Technology campus showing the \textit{KiQQer} project configuration. Alice, Bob (A and B in the main text, respectively) are connected -- via entanglement -- through a central node, Charlie (C). The quantum channel and associated classical communication links use hardware and software from: TNO (optical heads and quantum state analyzers); Qunnect (quantum networking infrastructure, including a bichromatic entangled photon-pair source); Single Quantum (cryogenically cooled superconducting nanowire single-photon detectors, SNSPDs); Qblox (quantum time modules); OPNT (White Rabbit synchronization modules); Fortaegis (physically unclonable function-based encryption hardware); and Xairos (quantum time-transfer software).}
\label{fig:experimental_architecture}
\end{figure*}

Robust quantum networks form the foundation of distributed quantum computing \cite{kimble2008quantum}. These networks depend on the efficient generation, transmission, storage, and measurement of entanglement. Long-distance entanglement distribution remains a major hurdle, as it demands high-performance quantum repeaters that are still under active development \cite{wehner2018quantum}. Meanwhile, targeted applications such as quantum key distribution (QKD) have accelerated the deployment of quantum communication infrastructure.

Moving from ``QKD-only" networks to those supporting broader applications involves substantial technical and integration challenges. Bridging this gap will require adapting entangled photon sources and quantum memories to serve both current market-driven applications, including QKD, secure communications, precision timing, and emerging quantum networking use cases. One critical step is addressing the mismatch between entanglement sources and quantum memories: high-brightness, broadband sources used in entanglement-based QKD~\cite{merolla2023high} are generally incompatible with quantum memories, which rely on narrow-band optical transitions \cite{askarani2021long, duranti2024efficient, wang2022field}. Recent results from platforms such as warm atomic rubidium vapors \cite{craddock2024high} and rare-earth-ion-doped crystals \cite{lago2023long} take a step towards bridging the source/memory gap with pair generation rates. Another challenge lies in managing hybrid transmission media for quantum signals. Optical fiber remains the preferred medium in urban environments due to its reliability and established infrastructure, while free-space links \cite{krvzivc2023towards}, including satellite connections, are essential in regions without fiber coverage or for mobile platforms \cite{bakker2024best}.

The \textit{KiQQer} experiment, named after Max Velthuijs' children’s book character \textit{Kikker}, celebrated for tackling complex challenges with honesty and collaboration, demonstrates entanglement distribution across both fiber and free-space channels. Built entirely from commercially available components within the Dutch Quantum Ecosystem, the deployed three-node network features a central node equipped with an entangled photon-pair source, connected to two receivers via a hybrid link. This architecture is relevant for scenarios such as connections to moving platforms, including aircraft, drones, and ships, as well as last-mile links \cite{Sheridan2025}. In the longer term, ground-to-satellite connectivity is a key application, with optical uplinks requiring significant advances to enable efficient entanglement distribution, a challenge that research institutes have been actively addressing in recent years \cite{vallone2015experimental, yin2017satellite}.

This demonstration marks a step forward in the technological maturity of quantum systems and confirms their potential compatibility with first-generation quantum network applications, including entanglement-based QKD \cite{bennett1992quantum}, quantum time transfer \cite{7}, and controlled physically unclonable functions \cite{vskoric2010flowchart}.

\section{Consortium and Network Architecture}~\label{sec:DataComm}
The \textit{KiQQer} consortium brought together complementary expertise from multiple partners, each contributing critical components to the demonstration. The field test was conducted over six weeks between August and Oct. 2024, with integration carried out at TNO and TU-Delft facilities in the presence of all partners. Table~\ref{tab:kiqqer_partners} summarizes each partner’s role in the experiment.

\begin{table}[h!]
\centering
\footnotesize 
\renewcommand{\arraystretch}{1.5} 
\setlength{\tabcolsep}{2pt} 
\rowcolors{2}{blue!20}{green!20} 
\begin{tabular}{p{2.5cm} p{5.5cm}}
\rowcolor{white}
\textbf{Partner} & \textbf{Contribution} \\
\hline
\rowcolor{gray!10}
\textbf{TNO} & Optical heads for free-space optical communications; integration of all subsystems. \\
\rowcolor{red!15}
\textbf{Qunnect} & Bichromatic source of polarization-entangled photonic qubit pairs. \\
\rowcolor{blue!15}
\textbf{OPNT} & White Rabbit synchronization modules. \\
\rowcolor{red!15}
\textbf{SingleQuantum} & 4-channel SNSPD-based detection system. \\
\rowcolor{blue!15}
\textbf{Qblox} & Qubit Timetag Modules compatible with the Qblox Cluster platform. \\
\rowcolor{red!15}
\textbf{Xairos} & Quantum Time Transfer protocol. \\
\rowcolor{green!10}
\textbf{Fortaegis} & Controlled Physically Unclonable Function encryption protocol for authentication of nodes. \\
\hline
\end{tabular}
\caption{\textit{KiQQer} consortium partners and contributions.}
\label{tab:kiqqer_partners}
\end{table}

\subsection{Block Diagram and Interface Definition}
\begin{figure}[h!]
\centering
\includegraphics[width=\linewidth]{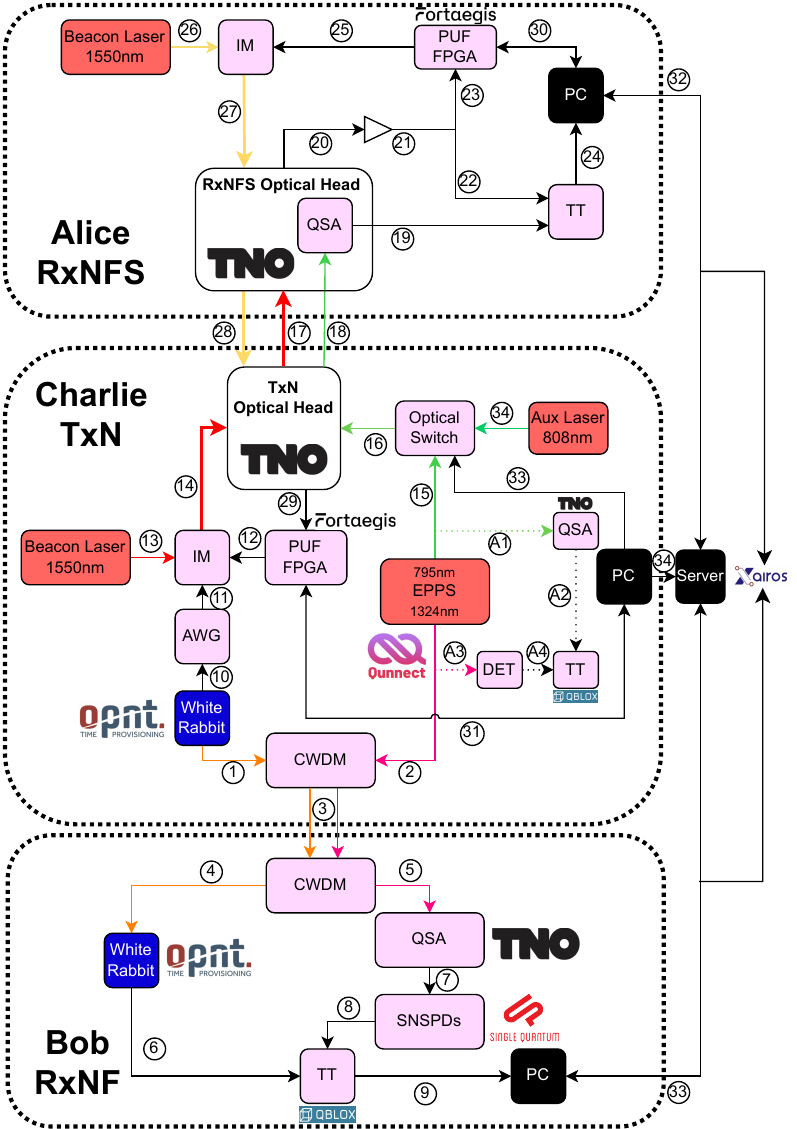}
\caption{Block diagram of the \textit{KiQQer} experimental setup showing system interfaces and hardware contributions. Partner company logos indicate the origin of each major hardware component used in the implementation.}
\label{fig:blockDiagramInterfaces}
\end{figure}

The \textit{KiQQer} experimental setup integrates both fiber and free-space optical channels, along with a diverse set of components that must interface precisely for successful operation. Figure~\ref{fig:blockDiagramInterfaces} presents the complete block diagram of the setup. Although all components are shown, some were used only during specific phases of data acquisition. For instance, the quantum state analyzer (QSA) at Node~C was inactive when photons were routed to Bob instead of to it. Where applicable, such distinctions are clarified in the text.

The \textit{KiQQer} network is composed of two edge nodes (A and B) interconnected via a central node (C). Node~C hosts a bichromatic polarization-entangled photon-pair source, producing signal photons at 795~nm and idler photons at 1324~nm. Both are spectrally isolated from the C-band, allowing low-loss multiplexing with classical communication signals. For example, at the transmitter node (TxN), the Master White Rabbit (WR) switch sends a 1000Base-LX Gigabit Ethernet (GbE) signal~(1) over single-mode fiber (SMF), multiplexed with idler photonic qubits~(2) via a coarse wavelength division multiplexer (CWDM). The combined signal~(3) is transmitted over fiber to the receiver node fiber interface (RxNF), where the CWDM separates the classical GbE~(4) from the quantum idler signal~(5).

At Node B, the RxNF WR switch recovers a 10~MHz clock and 1~PPS signal~(6) from the GbE stream and routes them to the local time tagger. The idler photon~(5) enters the QSA, where a sequence of waveplates defines its polarization measurement basis. A polarization beam-splitter (PBS) separates the selected (e.g., H) and orthogonal (e.g., V) states, and the four QSA outputs are detected by superconducting nanowire single-photon detectors (SNSPDs)~(7). Each detection generates a TTL pulse sent to the time tagger~(8), which records time-stamped data~(9) and streams it to the RxNF computer.

At Node C, the WR switch at the TxN also provides 10~MHz and 1~PPS references to an arbitrary waveform generator (AWG)~(10). The AWG outputs a 5~MHz clock with periodic start-sequence patterns~(11), phase-locked to the WR reference. Clock signals are combined with cPUF-encrypted FPGA data~(12) via Manchester encoding and amplified to drive an intensity modulator coupled to a beacon laser~(13). The modulated optical signal is delivered to the TxN optical head over SMF. Within the optical head, quantum~(16) and classical~(14) channels are fiber-to-free-space coupled and co-aligned for free-space transmission. The quantum channel originates from the entangled photon-pair source (EPPS)~(15), and the classical channel may optionally carry light from an auxiliary laser~(34) through an optical switch. These channels are launched over free-space toward the receiver node free-space interface (RxNFS)~(17,18). 

At Node A, the RxNFS optical head, a dichroic mirror separates the incoming beam. The classical component is used for pointing stabilization and clock/data recovery~(20), while the quantum signal is directed to the QSA. A free-space switching mirror enables redirection of the optical beam to a polarimeter when polarization alignment checks are required. In this configuration, the auxiliary laser at the TxN is selected by activating the optical switch~(16,34). The classical channel~(14,20) is converted to an electrical signal by a free-space–coupled avalanche photodiode (APD)~(20), amplified~(21), and split using an RF splitter~(22,23). One branch is used by the time tagger~(22) to align measurements with the clock and start sequence, while the other is interpreted by the cPUF FPGA~(23) for encrypted message decoding. For bidirectional communication, the RxNFS also sends a modulated beacon signal, which is received, demodulated, and decoded by the TxN using the same procedure~(25--29). Meanwhile, the QSA at RxNFS detects the signal photons~(18) and sends TTL pulses to the time tagger~(19), which forwards the recorded data~(24) to the local computer. Both the TxN and RxNFS computers interface with their respective cPUF FPGAs~(30,31), and communicate with TNO’s central servers~(32--34) over a secured VPN. Additional auxiliary interfaces (A1--A4) support future system integration. For clarity, internal links such as the EPPS control interface are omitted from the diagram.

\subsection{Node Deployment and Link Establishment}
Although the synchronization paths differ — fiber-based WR for Node~B and free-space pilot tone for Node~A — they are kept in phase via a common 10~MHz reference. This maintains global timing alignment across the entire system, as detailed in Figure 7. The signal photon is directed to Node~A through a $\sim$150~m long free-space link, originating from a spectrally diverse optical head at Node~C. Classical synchronization is achieved using a C-band pilot tone amplitude-modulated with the clock signal. The optical head includes a 150~mm aperture beam expander and active pointing control. This system uses the 1540~nm pilot tone received from Node~A, with 30\% of its power sent to a quadrant cell detector for tip/tilt feedback to a fine steering mirror (FSM). Details of this mechanism are discussed in the Appendix. The idler photon is routed through a CWDM and combined with a WR GbE signal. The composite signal travels over 200~m of fiber to Node~B. Figure~\ref{fig:experimental_architecture} shows the layout of the nodes at the TU-Delft campus.

At Node~A, the incoming free-space signal is collected by a 40~mm telescope. A dichroic mirror separates the quantum and classical signals. The classical signal path includes polarization separation for simultaneous bidirectional pilot tone handling. As in Node~C, 30\% of the inbound classical beam is used for pointing stabilization, while the rest is detected by a free-space APD for clock recovery. Quantum state analysis is performed at the receiver optical head. A symmetric beam-splitter implements passive basis selection, with one output path including a motorized half-wave plate to switch measurement bases. Both outputs go through PBSs, and the resulting four ports are directed to Si-APDs. These are connected to a time tagger synchronized with the recovered clock. The total optical loss from Node~C to A is 14.8~dB $\pm$ 0.2~dB, composed of geometric (10.4~dB $\pm$ 0.2~dB), window propagation (2.2~dB $\pm$ 0.1~dB), and internal reflection/transmission losses (1.2~dB $\pm$ 0.1~dB). The free-space synchronization protocol introduces 200~ps jitter.

Node~B separates classical and quantum signals using the same CWDM as Node~C. Clock synchronization is provided by WR modules, which output both clock and 1~PPS signals. These are used by the time tagger to time-stamp and batch detection events. The idler photons are analyzed in a fiber-coupled QSA identical in design to the one at Node~A but operating at 1310~nm. The four output paths of the QSA at Node~B are coupled into single-mode fibers and routed to SNSPDs for detection. Timing analysis is performed in the same manner as at Node~A: a time tagger receives the clock, 1~PPS, and detection signals from the single-photon detectors. Although the synchronization scheme between Node~C and Node~B (WR over fiber) differs from the one used between Node~C and Node~A (direct modulation of the clock onto the pilot tone over free-space), synchronism between detections at Nodes~B and~A is maintained by feeding the arbitrary waveform generator (AWG) with the 10~MHz reference signal output from the WR module. See Figure~\ref{fig:blockDiagramInterfaces} for further details on clock distribution and synchronization paths. At Node~C, because the InGaAs detectors are operated outside their optimal efficiency range, their performance is off specification and, at best, reaches $\sim$20\% efficiency.\\

\noindent The experiment proceeds in four stages: 
\begin{enumerate}
 \item \textbf{Optical path establishment.} This step is straightforward for the fiber path. For free-space, a coarse alignment is performed using a periscopic mirror at the transmitter.
 \item \textbf{Classical link establishment.} In fiber, the GbE signal is directly demultiplexed. In free-space, fine-pointing feedback is activated to maintain alignment. Synchronization is validated by comparing clock edges.
 \item \textbf{Coincidence peak detection.} The source at Node~C is verified through correlations with Nodes~A and B.
 \item \textbf{Entanglement visibility scans.} With delays established, visibility is evaluated by rotating the half-wave plates and extracting correlation data from coincidence counts.
\end{enumerate}

\section{Experimental Results and Analysis}
\begin{figure}[h!]
\centering
\includegraphics[width=\linewidth]{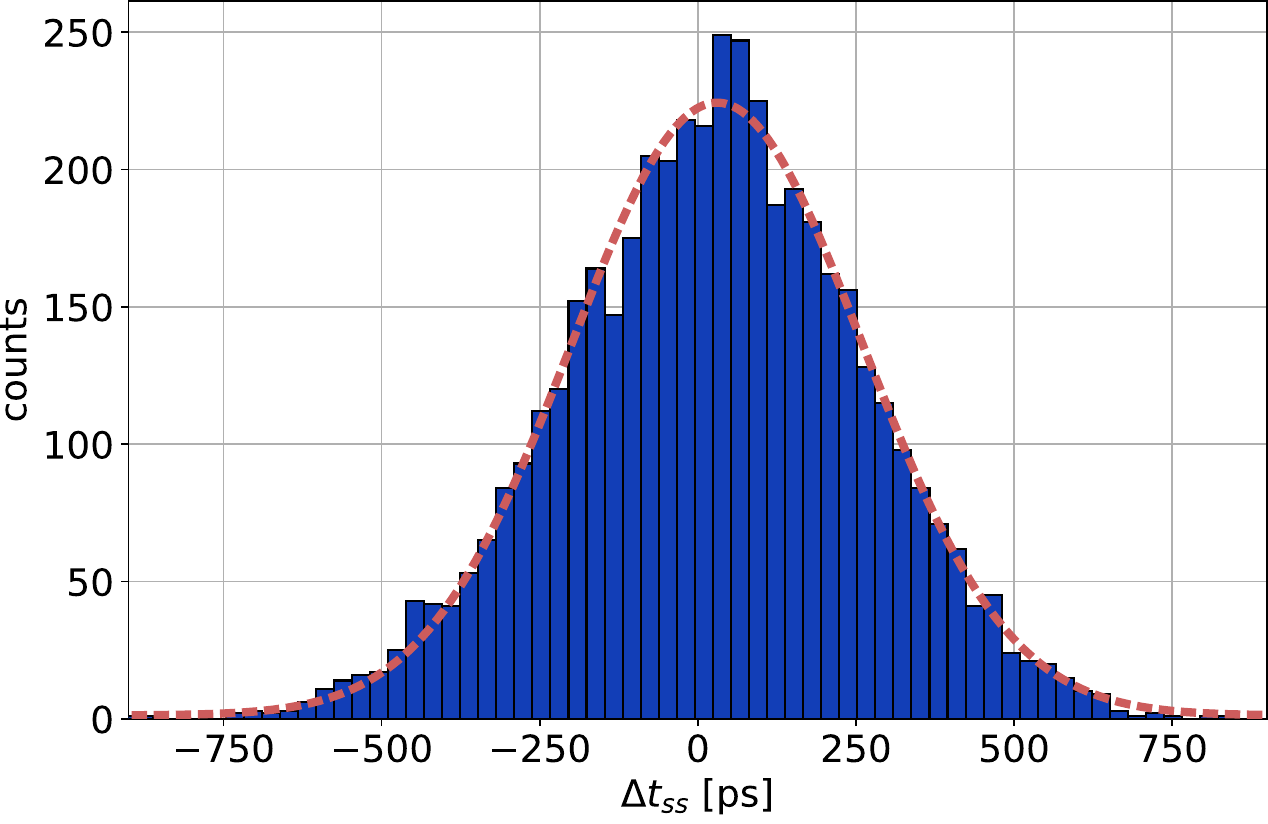}
\caption{Relative clock jitter between Nodes C and A. This measurement provides a lower bound on the time window spanned by the correlation peaks between the nodes. The orange line shows a Gaussian fit to the data, from which the peak’s Full Width at Half Maximum (FWHM) is estimated at 541~ps, representing the system jitter.}
\label{fig:Tx2RxJitter}
\end{figure}

The installation of the optical heads at Nodes C and A began with assembling tripods and coarse pointing alignment systems. A periscopic mirror was mounted in the transmitter optical head’s path to enable manual aiming, after which the transmitter pilot tone was turned ON. At the receiver optical head, a free-space optical power meter positioned at the focal plane of the optical relay was used to verify light collection. Manual adjustments were performed until power was detected at the receiver’s QuadCell, at which point the fine pointing alignment system was engaged. Once the receiver pilot tone was turned ON, additional manual adjustments were made before enabling fine pointing alignment on the transmitter side as well. Activation of the AWG modulating the transmitter pilot tone then allowed evaluation of the recovered clock signal at the receiver node. The alignment and verification process was completed within two days and enabled measurement of the synchronization system’s jitter between Nodes C and A, as shown in Fig.~\ref{fig:Tx2RxJitter}. Jitter results were obtained by evaluating the relative arrival times of the periodically transmitted start sequences embedded in the clock signal and generating a histogram of the events. The correlation measurements revealed an initial 150~ns time delay, corresponding to the static relative delay between Nodes C and A due to link distance and electronics latency. This static delay was compensated in software for all results to be centered at $\Delta t = 0$. All data presented in this section were acquired using an external time tagger (TimeTagger20, Swabian Instruments) at each node; although the integration window used to evaluate correlation peaks may change due to the achievable timing jitter of the setup, the time bin is kept constant at 100~ps. Tests using the Qblox Qubit Timetag Module\footnote{Due to the limited availability of the Beta module during the free-space trials, Qblox hardware tests were restricted to the Bob and Charlie nodes.} are detailed in the Appendix.

The measurement of Figure \ref{fig:Tx2RxJitter} is foundational to the following correlation measurements between entangled photons, both for synchronization and for data acquisition and processing. Synchronization via clock recovery at the receiver node includes detection of a start sequence periodically embedded in the signal. The time tagger is programmed to identify this pattern and implement a phase-locked loop between the received clock and its internal reference. This enables (i) batching of detection events so that Nodes~C and A share a common temporal reference for data correlation, and (ii) maintaining a stable and common time-base derived from the same clock source.

\begin{figure}[h!]
\centering
\includegraphics[width=\linewidth]{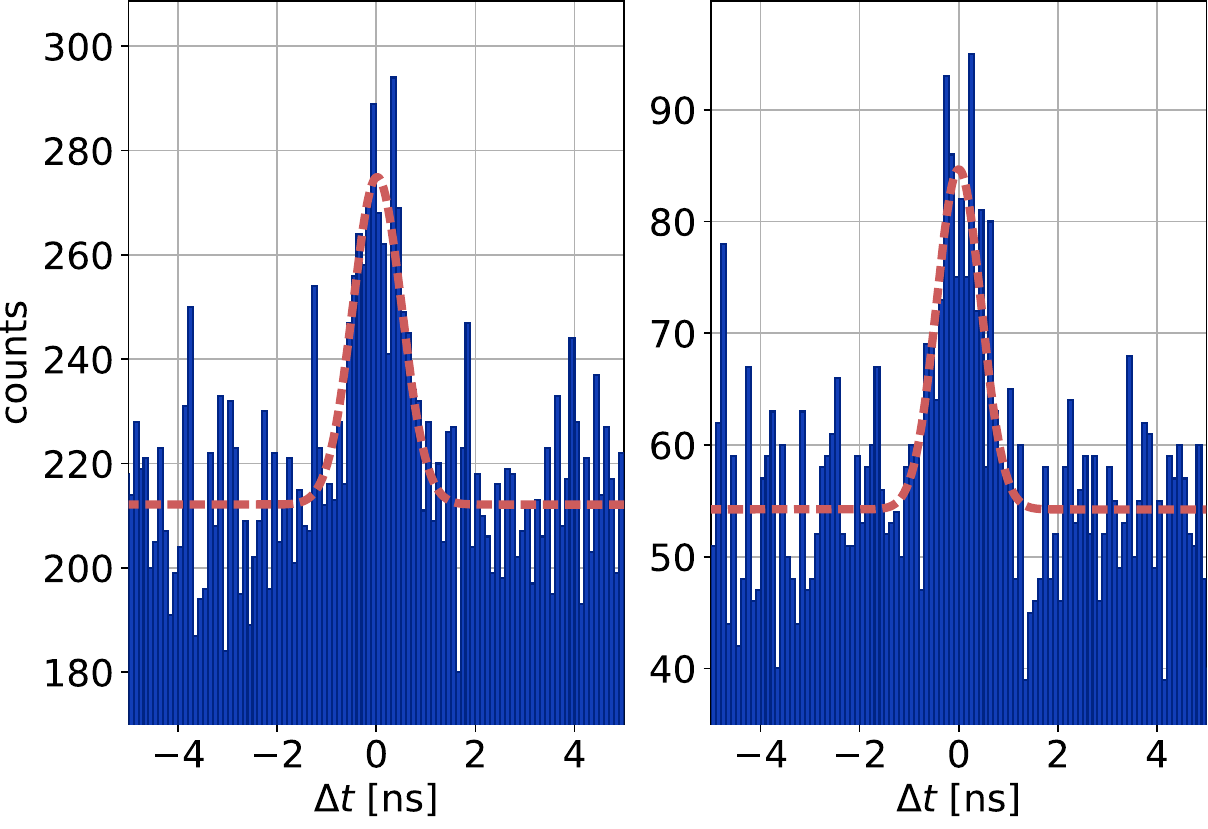}
\caption{Photon correlation measurements recorded in daylight on 27~Sept.~2024, using the synchronization, data acquisition, and processing methods described in the text. Data were accumulated over a 60~s interval, measuring coincidences between two Si-APDs at Node~A and the InGaAs APD at Node~C. The CAR is noise-limited, not source-limited.}
\label{fig:dayTimePeak}
\end{figure}

Batch-based processing significantly accelerates the correlation analysis, since the algorithm scales with the product of the dataset sizes from each node. During operation, batches are generated once per second, resulting in a corresponding one-second processing interval. Each measurement run is assigned a unique identifier (UUID), which is stored in a shared database hosted on a server connected to the TNO VPN. Node~C’s control computer retrieves the UUIDs as soon as they appear in the database and proceeds with batch processing. Using this procedure, and with the entangled photon-pair source connected to the qubit path of the free-space channel, the correlation measurements of Fig.~\ref{fig:dayTimePeak} were recorded in daylight on 27~Sept.~2024. The figure of merit for these initial results is the Coincidence-to-Accidental Ratio (CAR), which serves as an approximation to the second-order cross-correlation function at zero delay, $g^{(2)}_{1,2}(0)$.

Although the results in Figure~\ref{fig:dayTimePeak} are not sufficient to demonstrate entanglement between the two nodes, they show that such measurements are feasible during daytime, provided that spectral filtering is improved. The band-pass optical filter installed at Node~A has a bandwidth of 2~nm (approximately 188~GHz at the 795~nm operating wavelength), whereas the photonic qubits have an estimated bandwidth of only 1~GHz. Narrowing the filter bandwidth could substantially reduce noise levels, which during daytime exceed 1~Mcps at the Si-APDs. Such high noise rates not only raise the accidental-coincidence background but also increase detector deadtime, thereby reducing the probability of registering true photon events.

Since the filtering of classical power and stray light from the pilot tones is less critical than for the single-photon detectors, it was possible in this configuration to operate the bidirectional free-space channel. Over this channel, data was encrypted using a cPUF combined with AES-256-GCM between Nodes~A and~C. Pre-characterized FPGAs were installed at both nodes and connected to the pilot-tone modulators, enabling the encrypted messages to be transmitted together with the clock signal via Manchester encoding. At the receiver, the signal was conditioned and decoded by the opposite FPGA, confirming that the established physical layer can support simultaneous classical communication and photonic-qubit transmission.

\begin{figure}[h!]
\centering
\includegraphics[width=\linewidth]{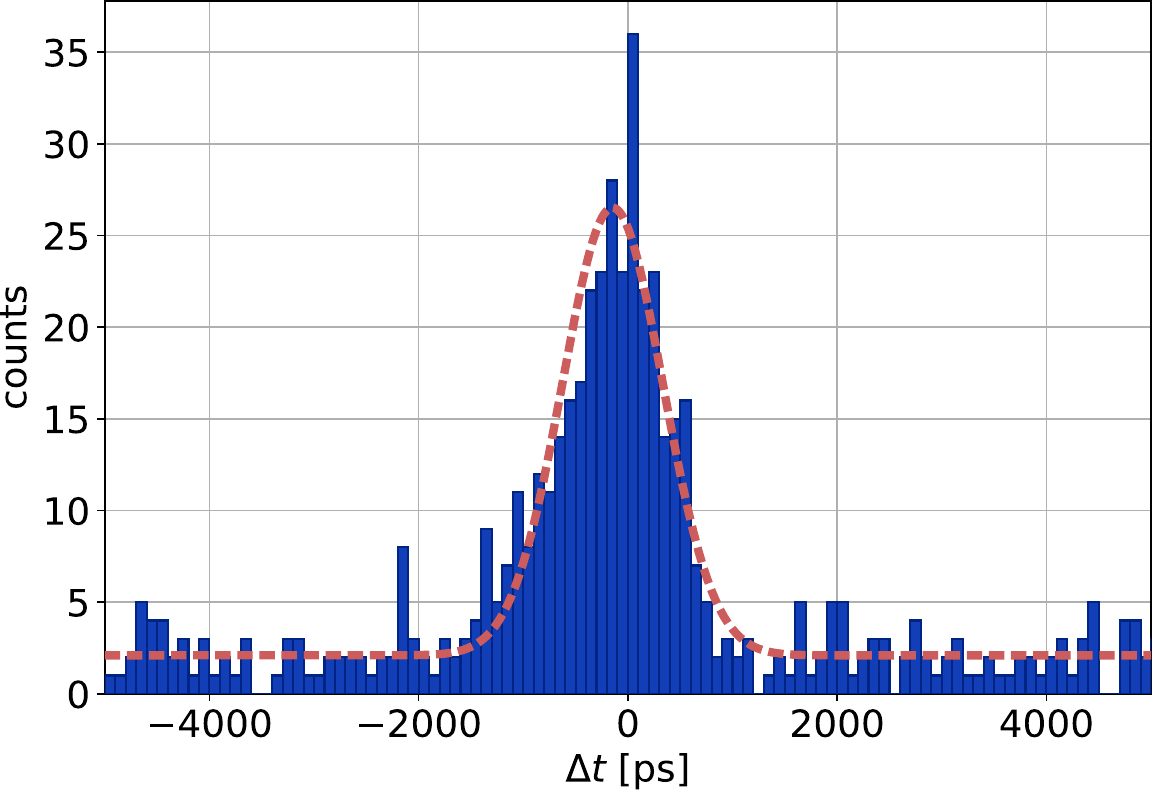}
\caption{First correlation measurements recorded at night on 2~Oct.~2024. The reduced noise level reveals the non-classical nature of the photon pairs detected at both nodes and yields a signal-to-noise ratio approaching that required for demonstrating entanglement distribution.}
\label{fig:nightTimePeak}
\end{figure}

Following daytime measurements, the correlation-peak experiments were repeated at night, when stray-light counts dropped to $\sim$ 4k~cps; close to the detector dark-count rate of 1k~cps. These results, acquired on 2~Oct.~2024 with the same experimental parameters as in Figure~\ref{fig:dayTimePeak}, are shown in Figure~\ref{fig:nightTimePeak}.

\begin{figure}[h!]
\centering
\includegraphics[width=\linewidth]{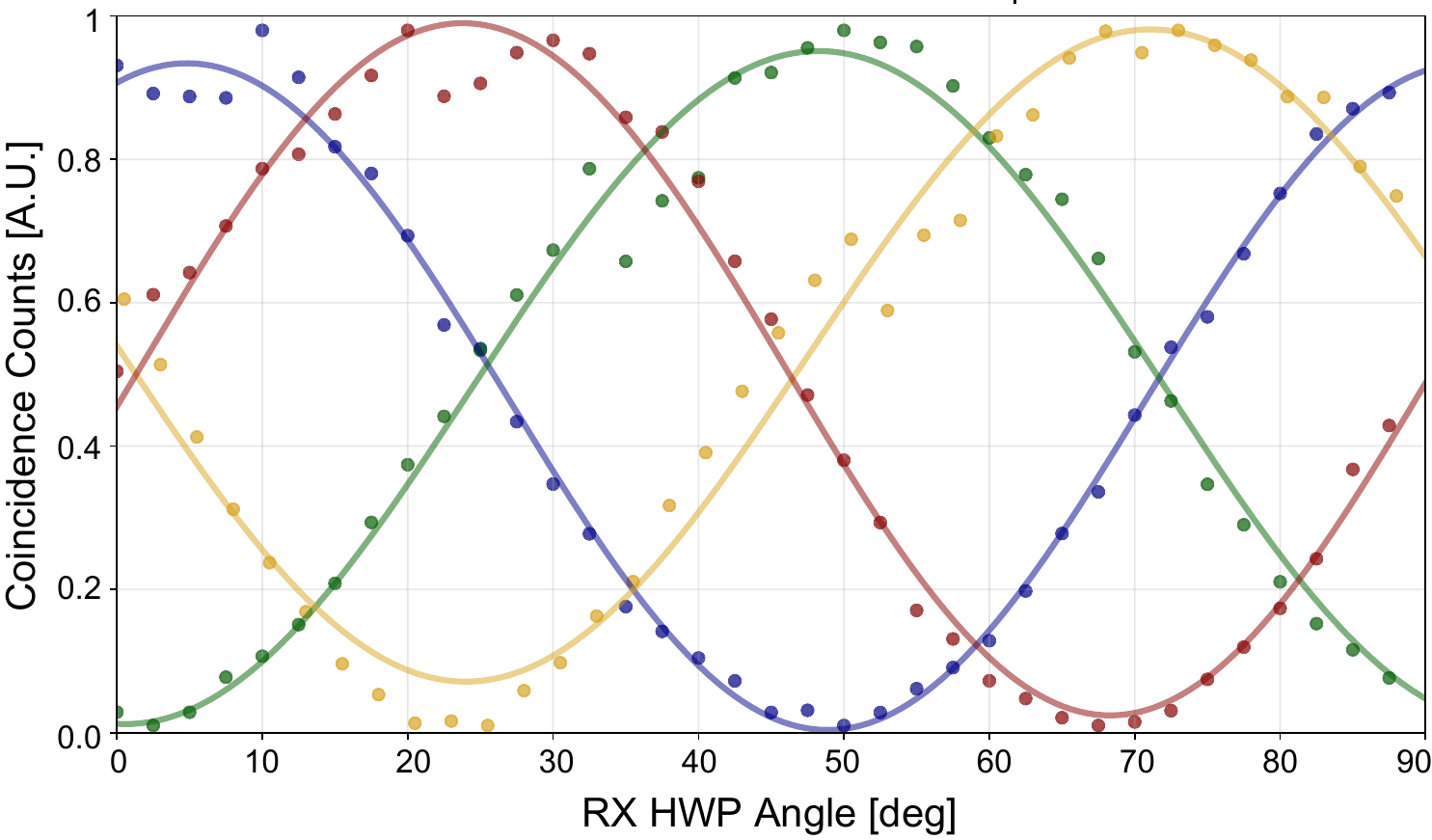}
\caption{CHSH entanglement visibility between Nodes~C-~B using the bi-chromatic entangled photon-pair source operated at 10-20\% of maximum brightness. Each point represents a 30~s integration and is taken over a selective region of interest (ROI). The sinusoidal fits correspond to the expected quantum correlations for optimal measurement settings. Measured visibilities exceed the classical bound, yielding $S = 2.55$, and confirm the non-classical nature of the photon pairs.}
\label{fig:Tx2fiberRx}
\end{figure}

The results in Figure~\ref{fig:nightTimePeak} prompted two parallel efforts in the following months: (i) optimization of the free-space channel between Nodes~C and~A, including improved coarse pointing alignment and adjustment of beam divergence, and (ii) extraction of correlation measurements between Nodes~C and~B. During the free-space optimization, a mechanical instability in the transmitter optical head was identified as the enclosure lid introduced torsion on the optical breadboard, significantly affecting the relative pointing between the transmitter pilot tone and the photonic qubit optical paths. After correcting issues, the coupling efficiency between the two nodes over the free-space link improved noticeably. For the C–B link, the more stable fiber channel, combined with the ability to operate efficiently during daytime, enabled timely extraction of the entanglement visibility curves shown in Figure~\ref{fig:Tx2fiberRx} (8~Oct.~2024). Apart from distributing the clock and 1~PPS signals over fiber using WR modules, acquisition and processing followed the same protocol as for the C–A measurements. The SNSPDs in the laboratory, which offer lower jitter (13~ps), higher efficiency (91\% at 1324~nm), shorter deadtime (30~ns), and lower dark count rates ($\leq$2~cps), provided a significantly higher signal-to-noise ratio for these measurements. The increased detection performance allowed the source brightness to be reduced compared with previous runs, improving the fidelity of the generated quantum states. The FWHM of the correlation peaks, averaged over all measurements, was $995 \pm 87$~ps; accordingly, the integration window was set to 1~ns to maximize the signal-to-noise ratio.

The relevant figures of merit to extract from Figure~\ref{fig:Tx2fiberRx} are: an average visibility of 94.3\%, an average QBER (H/V and D/A) of 2.8\%, an overall detection rate for the target channel combinations (HV, VH, DA, AD) of 344~s$^{-1}$, and an $S$-value of 2.55. The corresponding secret key rate in this configuration, taking into account a key extraction efficiency factor of 0.56 \cite{attema2021optimizing} consistent with the measured QBER and block length, privacy amplification, and error correction protocols, is 770~bits/s. This performance serves as a reference for the final entanglement distribution measurements between Nodes~A and~B (Figure~\ref{fig:Rx2RxEntVis}), carried out during the night of 11~Oct.~2024. 

\begin{figure}[h!]
\centering
\includegraphics[width=\linewidth]{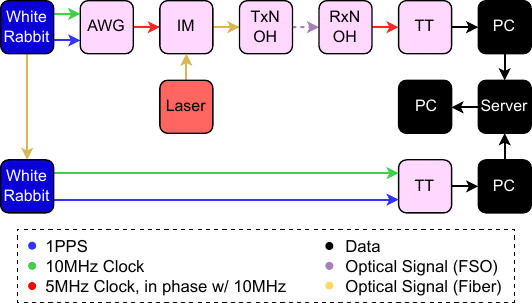}
\caption{Experimental configuration for synchronizing Nodes~A and~B via Node~C. The White~Rabbit module’s clock signal is used to lock the AWG’s internal clock, while the 1~PPS output serves as a trigger to periodically generate the start-sequence waveform at a matching rate.}
\label{fig:Rx2RxSynch}
\end{figure}

\begin{figure}[h!]
\centering
\includegraphics[width=\linewidth]{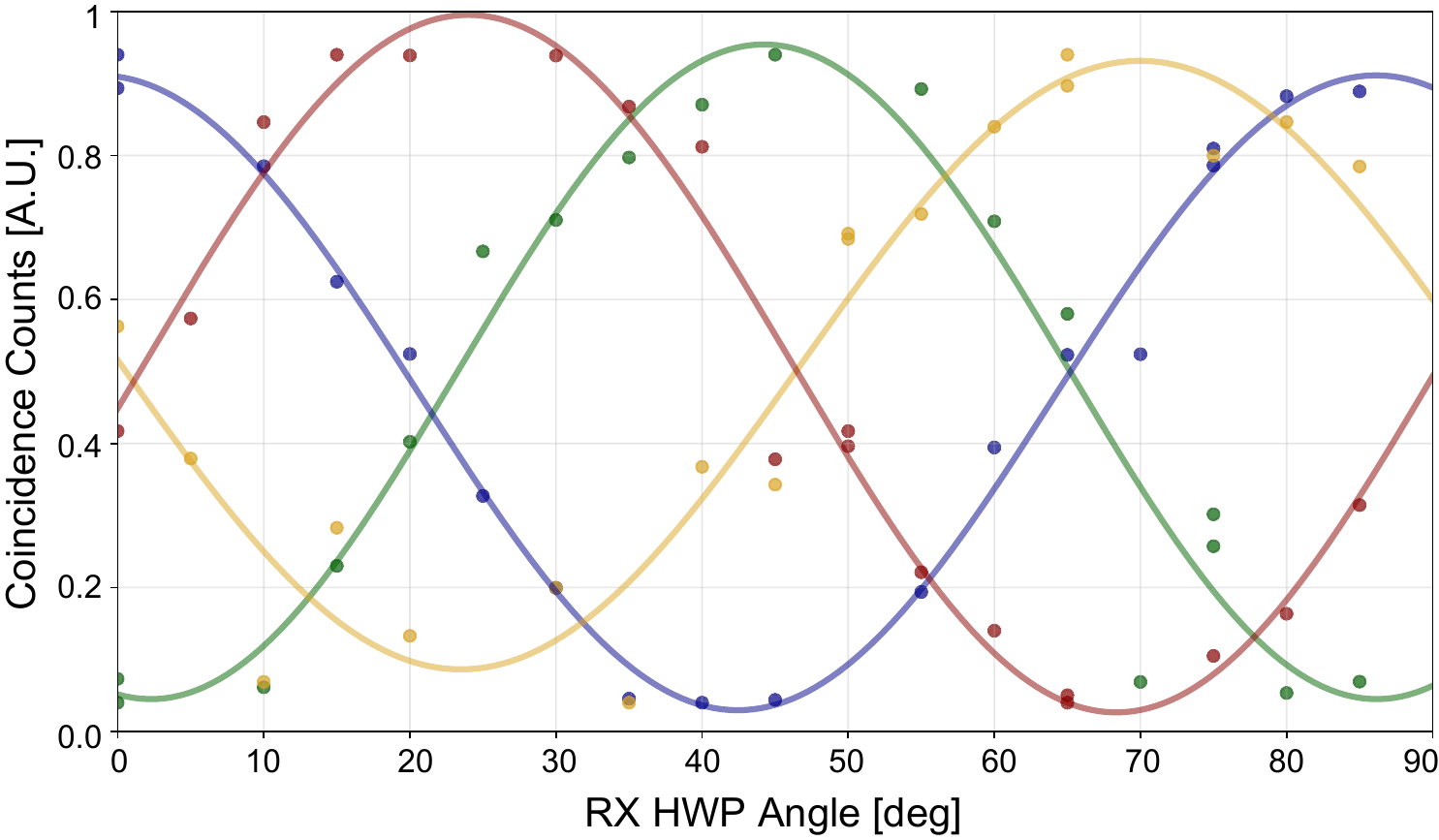}
\caption{CHSH visibility between Nodes~A-~B, with each point representing a 30~s integration of the correlation peak - over a narrow ROI. Polarization bases at both nodes remained matched over the 2~h acquisition, enabling stable curve extraction and yielding $S = 2.63$, above the classical bound.}
\label{fig:Rx2RxEntVis}
\end{figure}

In this configuration, mutual synchronization between nodes was achieved by using the White~Rabbit clock and 1~PPS signal from Node~C as a reference for the AWG generating the synchronization signals transmitted over free-space to Node~A (Figure~\ref{fig:Rx2RxSynch}). A detailed overview of the interfaces in the experimental setup is provided in Figure~\ref{fig:blockDiagramInterfaces}. Due to hardware limitations, remapping the White~Rabbit 10~MHz clock reference to the 5~MHz output of the AWG was necessary to prevent saturation of the time tagger’s communication bandwidth at the receiver.

The measured FWHM of the correlation peaks between the two edge nodes is $1077 \pm 106$~ps, higher than that observed between Nodes~C and~B, as expected from the additional clock jitter introduced by the free-space synchronization protocol. Accordingly, the correlation peak integration window was set to 1.1~ns, i.e., events within a $\pm550ns$ from the zero relative time delay were considered for the calculation of rates and fidelity. The results in Fig.~\ref{fig:Rx2RxEntVis} confirm successful entanglement distribution in the three-node configuration. Compared with Fig.~\ref{fig:Tx2fiberRx}, the free-space optical channel introduces additional losses, reducing the total detection rate to an average of 15~counts/s per channel combination. With an average QBER across the two measurement bases of 2.88\% (corresponding S-value of 2.63) and a key extraction efficiency of 0.56, the achievable secret key rate is 33~bits/s. The reduction from 770~bits/s for Nodes~C--B to 33~bits/s for Nodes~A--B (13.6~dB$\pm$ 0.2~dB) agrees with the estimated optical losses between Node~C’s optical head and Node~A’s QSA (13.8~dB$\pm$ 0.2~dB), as discussed in the previous section. Each data point in Fig.~\ref{fig:Rx2RxEntVis} corresponds to a correlation peak accumulated over 30~s; however, data upload and processing constraints prevented continuous acquisition. In addition, synchronization errors of timetaggers caused some data sets to be discarded, resulting in a total measurement time of 2~h. Within this period, the alignment between the QSAs at Nodes~A and~B remained sufficiently stable to allow extraction of results without intermittent manual adjustments.

Further optimization of the experimental setup was feasible but could not be carried out due to limited access to the locations of Nodes~C and~A. The results shown in Figure~\ref{fig:Rx2RxEntVis}, obtained in the early hours of 12~Oct.~2024, mark the conclusion of the \textit{KiQQer} experimental campaign and demonstrate successful hybrid fiber/free-space entanglement distribution between two edge nodes, mediated by a central node.

\section{Discussion}
Beyond demonstrating entanglement distribution with off-the-shelf hardware integrated by the partners, the goal is to connect the demonstration to concrete future use cases. In addition to QKD, which the setup has already shown it could support, \textit{KiQQer} targets performance levels compatible with cPUF-based classical encryption and quantum time transfer.

\subsubsection{Encryption and Physically Unclonable Functions}
With a secret key generation rate of 33~bits/s and the maximum recommended transfer size for AES-256-GCM of $2^{39}-256$~bits \cite{nist2007sp} within the cPUF, the corresponding maximum achievable data throughput over the data channel is $\sim$60~Gbit/s. In other words, the necessary measurement time until a new 256-bit-long seed is generated over the QKD channel is seven seconds; with a new seed, $\sim$500~Gbits of data can be encrypted, leading to a data rate of $\sim$60~Gbit/s. The successful transmission of cPUF-encrypted data alongside photonic qubits over the free-space channel demonstrates that such a rate could, in principle, be achieved between Nodes~A and~C. In the present demonstration, the data transfer was operated at a much lower rate, consistent with Manchester encoding of a 5~MHz clock signal. Nonetheless, extending the receiver optical head to support additional wavelength-multiplexed classical channels represents a straightforward upgrade path. In particular, replacing the free-space-coupled classical detector with fiber coupling would enable the use of standard dense wavelength division multiplexing (DWDM) equipment. The spectral filtering in the photonic qubit channel (795~nm) already permitted separation of a classical beam carrying 20~mW (13~dBm) at the transmitter and 0.5~mW (-3~dBm) at the receiver. This optical power budget is sufficient to support data rates exceeding 100~Gbit/s. Any coupling losses introduced by fiber integration could be mitigated by increasing the transmitter power beyond 20~mW, which in this experiment was limited solely by eye-safety regulations.

\subsubsection{Quantum Time Transfer}
In environments where Global Navigation Satellite System (GNSS) signals are unavailable or untrustworthy due to jamming, spoofing, or mission-specific constraints, secure and precise timing remains essential. Quantum Time Transfer (QTT) provides a means of achieving synchronization without relying on existing RF-based satellite infrastructure, while offering potential advantages in tamper resistance and resilience. Many operational scenarios impose strict size, weight, and power (SWaP) limits, restricted communication bandwidth, and short acquisition windows. Under such conditions, timing systems must deliver high accuracy, rapid lock-in, and robustness to environmental variations. QTT systems could address these demands, especially where conventional timing infrastructure is unreliable or inaccessible.

QTT protocols typically involve the bidirectional distribution of entangled photon pairs across a network. One photon from each pair is transmitted to a remote node, while its partner is retained locally. Each node detects both its locally received photon and the photon kept on-site, time-stamping each event with its own internal clock. By correlating detection events between nodes, estimates of both the relative clock offset and the propagation delay can be extracted. This enables high-precision time synchronization between physically separated systems, even in the absence of GNSS references.

For both directions, the detection times ($t_B^{rx}$, $t_A^{rx}$) are modeled with respect to each node’s local clock functions. The reduced forms of these expressions are as follows.
\begin{align*}
\textup{Downlink:} \quad t_B^{rx} &= g(t_B) + \varepsilon\\
 &\approx g(t_A + T_{AB} + D_{AB}) \\
 t_A^{local} &= f(t_A) + \varepsilon. \\\\
\textup{Uplink:} \quad t_A^{rx} &= f(t_A') + \varepsilon\\
 &\approx f(t_B' + T_{AB} + D_{BA}) \\
 t_B^{local} &= g(t_B') + \varepsilon.
\end{align*}
Here, $f(\cdot)$ and $g(\cdot)$ are the clock functions at nodes A and B, $T_{AB}$ is the unbiased time-of-flight between the nodes, $t_{A/B}'$ and $t_{A/B}$ represent idealized correlated photon-pair timestamps at nodes A and B, respectively, $\varepsilon$ is the uncertainty due to the measurement apparatus, and $D_{AB/BA}$ represent any time-of-flight asymmetry (w.r.t. $T_{AB}$) from A to B, or vice-versa; this value is relevant when the nodes are moving relative to each other, but can be neglected in the current configuration of the experiment.

By computing the cross-correlation between the sets of time tags, two latencies are obtained: \( \tau_{DL} \) and \( \tau_{UL} \). These represent the peak positions of the downlink and uplink correlations, respectively. Their sum and difference yield the time-of-flight \( T_{AB} \) and the clock offset \( \delta \) between the nodes:
\begin{equation}
\tau_{DL} = \textup{MAX}\{C_{AB}\} \approx T_{AB} + \delta
\end{equation}
\begin{equation}
\tau_{UL} = \textup{MAX}\{C_{BA}\} \approx T_{AB} - \delta
\end{equation}

This approach follows the standard framework of two-way time transfer and is not exclusive to quantum implementations. In the QTT protocol used here, entangled photons, single-photon detectors, and coincidence-based correlation measurements replace classical optical pulses. The result is a quantum method for time transfer.

\begin{figure}[h!]
\centering
\includegraphics[width=\linewidth]{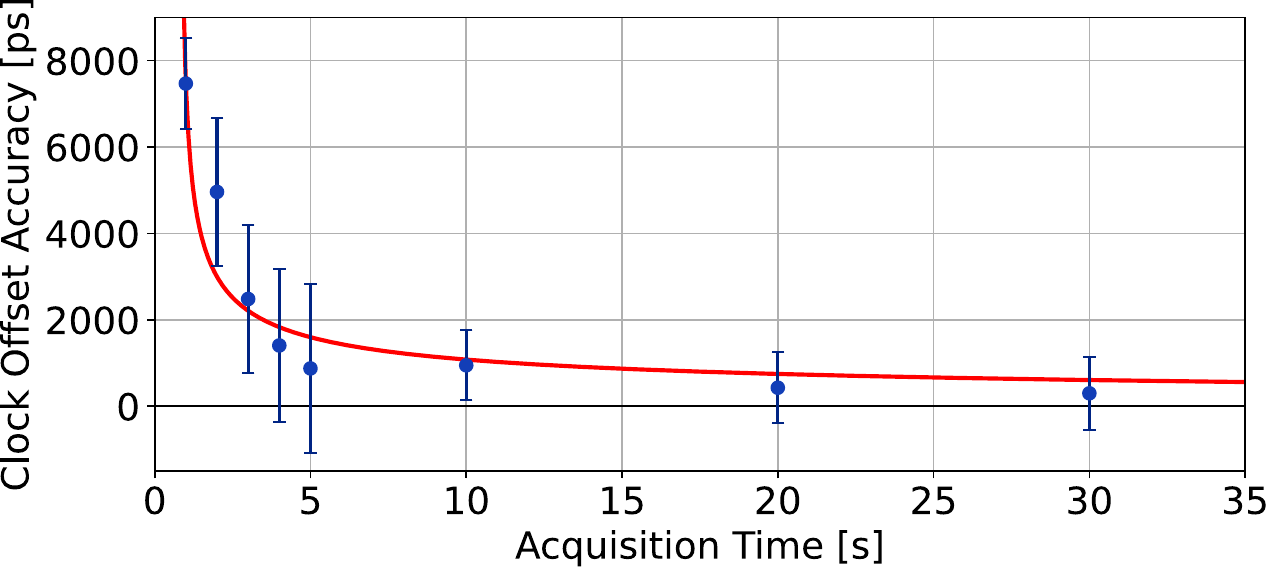}
\caption{Clock offset estimation using QTT algorithm and correlation data from the \textit{KiQQer} experiment. The fit corresponds to an inverse square root function ($\sim 1 / \sqrt{t}$). This is related to the intrinsic counting (Poisson) process of accumulating correlated detections, which improves the SNR ratio of the clock offset estimate according to the inverse square root. The fit tends asymptotically to the detector jitter.}
\label{fig:QTT}
\end{figure}

The data from this experiment were processed using Xairos’ QTT algorithm (Fig.~\ref{fig:QTT}). Although bidirectional operation is required to remove clock offset in real time, the analysis demonstrates consistent timing alignment. Short acquisition windows introduce uncertainty in the clock-offset estimate, largely due to limited photon counts and the presence of multiple propagation paths within the polarization-resolving QSA. By integrating over 30~s, the residual uncertainty was reduced to 594~ps, in line with the 500~ps jitter of the APDs. This jitter dominates the temporal noise budget. With a reciprocal QTT link, further improvements in clock offset resolution would be expected, thanks to the cancellation of clock drifts and better mitigation of path-dependent variations.

The results support the suitability of a \textit{KiQQer}-like architecture for quantum time transfer between mobile or distributed nodes in metropolitan environments, which also lays the foundation for more advanced applications such as secure positioning, where combining QTT with QKD protocols can protect time-of-arrival information from adversarial interference.

\section{Conclusion}
A key challenge for Research and Technology Organizations (RTOs) like TNO is bridging the TRL gap between promising scientific advances and operational technologies. In quantum technologies, this challenge is magnified by fragmented supply chains and commercial readiness. The \textit{KiQQer} project addresses this gap by demonstrating metropolitan-scale entanglement distribution over hybrid fiber and free-space links, relying exclusively on off-the-shelf components. In collaboration with Qunnect, Single Quantum, Qblox, OPNT, Fortaegis, and Xairos, TNO developed and field-tested a three-node network capable of distributing polarization-entangled photon pairs. Central to this system, the bichromatic entangled photon-pair source enabled hybrid-channel operation, seamlessly integrating fiber-based infrastructure with free-space optical links. The resulting architecture could support a range of applications, including QKD, QTT, and authentication via cPUFs.

This demonstration shows that practical quantum networking can extend beyond controlled laboratory environments. By validating real-world entanglement distribution with commercially available tools, \textit{KiQQer} establishes a tangible foundation for early quantum network deployments and opens the path toward untrusted-node and memory-enabled architectures.

\section{Acknowledgements}
The \textit{Metropolitan Free-Space Entanglement-based Quantum Key Distribution and Synchronization (KiQQer)} project is funded by Holland High Tech~|~TKI HSTM through the PPP Innovation Scheme (PPP-I) for public--private partnerships. G.~C.~A.\ thanks Bob Dirks, Charlotte Postma, Shobhit Yadav, and the Quantum in Space Roadmap of TNO for their contributions to the consortium’s formation and project execution. The TNO team gratefully acknowledges the colleagues and friends of the Quantum Technology Department for sharing office space to enable experimental data collection, as well as the infrastructure managers and technical staff of TNO Stieltjesweg. The team also thanks the TU Delft partners for providing the location of the free-space receiver node used in the experiment. 

\bibliography{sample}

\clearpage
\newpage

\section{Appendix}
\subsection{Spectrally Multiplexed Optical Free-Space Optical Heads}
\begin{figure}[h!]
 \centering
 \includegraphics[width=0.99\linewidth]{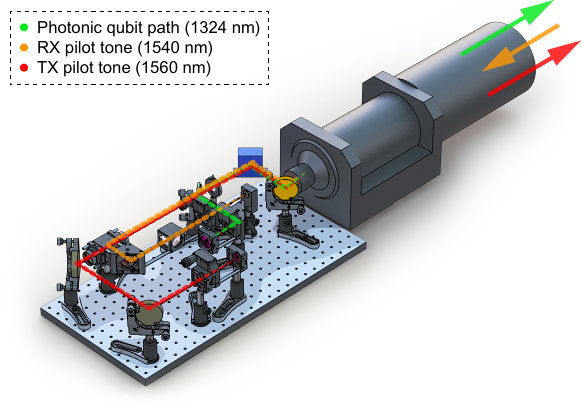}
 \caption{Transmitter breadboard. Three optical paths are present: the transmitter pilot tone (red), originating at the bottom left and routed clockwise around the breadboard; the photonic qubit path (green), starting at the center and exiting toward the beam expander (BEX) on the right; and the receiver pilot tone (orange), entering from the BEX and, after reflection by the PBS, directed to the QuadCell and the free-space–coupled detector.}
 \label{fig:txn}
\end{figure}

Node~C features a TNO custom-designed optical head, shown in Figure~\ref{fig:txn}, that manages three spectrally and polarization-multiplexed optical signals: (i) entangled photonic qubits, separated from the other channels via a dichroic mirror (DM); (ii) an outgoing classical pilot tone; and (iii) an incoming pilot tone. The two classical signals are separated by polarization using a polarizing beam-splitter (PBS). These paths are indicated in Figure~\ref{fig:txn} as red (qubit), orange (outgoing pilot tone), and green (incoming pilot tone). All beams are directed into a 10$\times$ beam expander (BEX) with a 150~mm aperture via a folding mirror. Immediately before the folding mirror, a fine-steering mirror (FSM) provides dynamic beam alignment.

The FSM is continuously actuated by a PID feedback loop that receives input from a quadrant-cell photodetector (QuadCell). This loop centers the incoming green beam on the QuadCell and, due to the symmetric reflection geometry, simultaneously aligns the outgoing pilot tone and qubit beams to the remote receiver’s aperture. The same loop ensures that the incoming pilot tone is directed accurately onto a free-space–coupled avalanche photodiode (APD), which handles downstream data transmission from Node~A to Node~C, as described in Section~\ref{sec:DataComm}.

\begin{figure}[t!]
\centering
\includegraphics[width=\linewidth]{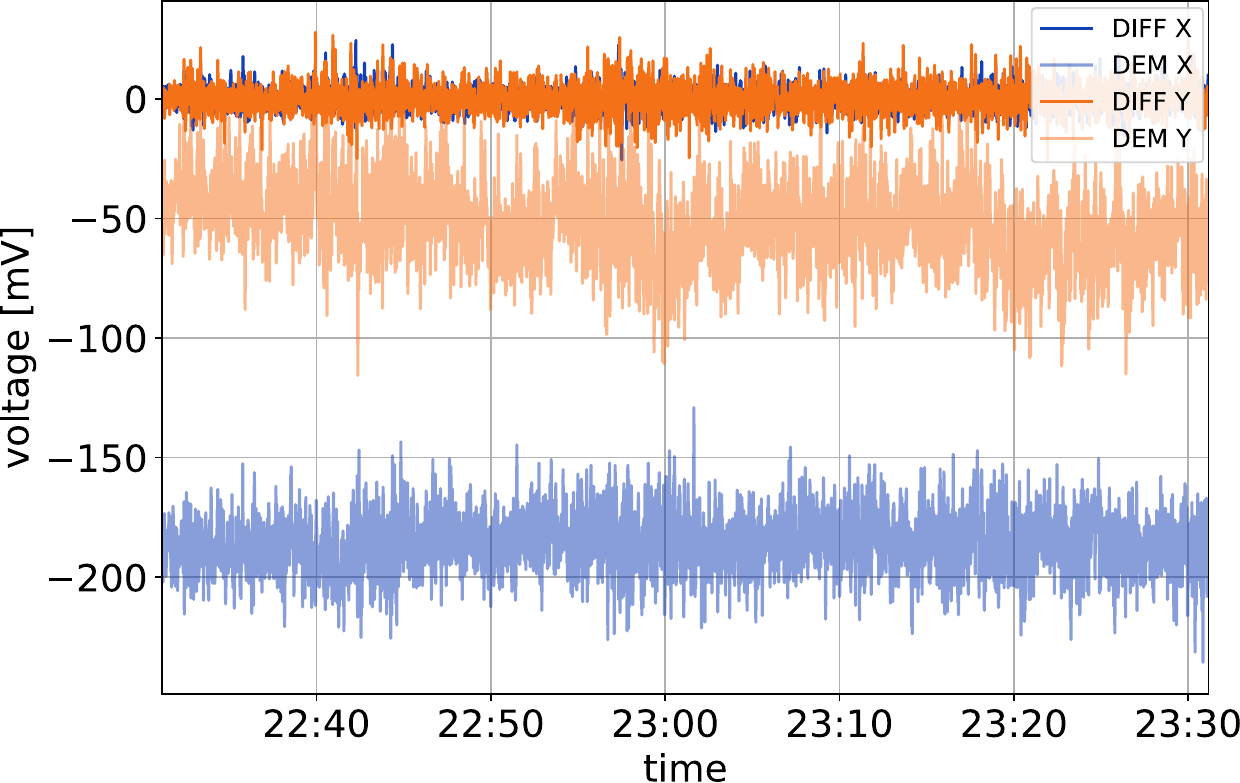}
\caption{Transmitter QuadCell data recorded over a one hour measurement. Plotted are the voltage differences measured for the horizontal (X) and vertical (Y) axes of the QuadCell (DIFF), along with the corresponding control voltages applied to the FSM in the transmitter optical head (DEM). Based on these signals, the pointing stability is estimated at $\sim$16~\textmu rad over the duration of the experiment; the calculation takes into account the beam diameter at the BEX entrance (80mm), the measured average optical power (0.6mW) and photodiode gain (1V/mW), the effective focal length towards the QuadCell (500mm), and the standard deviation of the measured signal in both directions (5mV).}
\label{fig:txQC}
\end{figure}
\begin{figure}[h!]
 \centering
 \includegraphics[width=\linewidth]{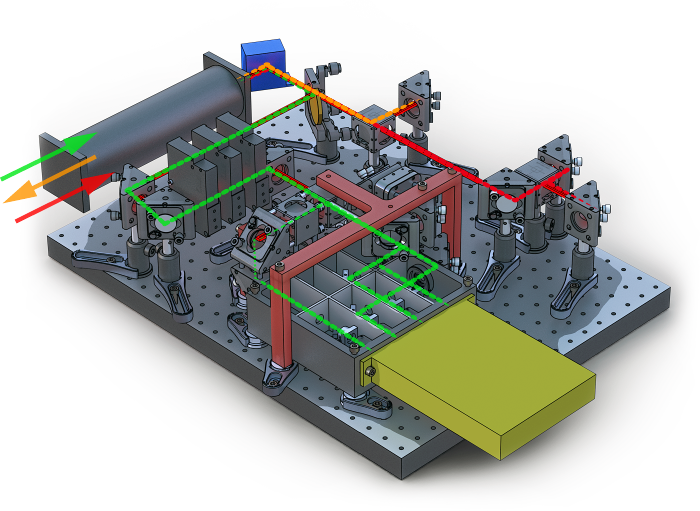}
\caption{Receiver breadboard. The setup contains three optical paths: the receiver pilot tone (orange), entering from the top left and reflected by a PBS toward the beam expander (BEX); the photonic qubit path (green), reflected by a dichroic mirror and directed into the Quantum State Analyzer for measurement; and the transmitter pilot tone (red), entering from the top right and transmitted through the PBS to the QuadCell and the free-space–coupled clock detector. The yellow-enclosure contains four Si-APDs for single-photon detection.}
 \label{fig:rx_path}
\end{figure}

Figure~\ref{fig:txQC} shows a one-hour continuous record of the pointing control signals at Node~C, demonstrating the long-term operational stability of the FSM–QuadCell feedback loop. The receiver breadboard shares core design features with the transmitter system but integrates a polarization-based quantum state analyzer (QSA) operating at 795~nm directly into the optical layout. This integration is enabled by free-space–coupled silicon-based single-photon detectors (SPDs), as depicted in Figure~\ref{fig:rx_path}.

The receiver breadboard, like its counterpart, manages three optical paths. The quantum and classical beams are spectrally separated, while the two counter-propagating classical tones are separated via polarization. Although integrating the QSA increases optical complexity, it eliminates fiber-coupling losses, improving the system’s quantum efficiency (see the lower-right optical path in Figure~\ref{fig:rx_path}). The RX breadboard is also responsible for clock recovery. The transmitter’s pilot tone is amplitude-modulated, and after spectral and polarization demultiplexing, the received signal is split along two paths: one for pointing feedback via the QuadCell and FSM, and the other for clock signal extraction via a free-space–coupled detector. While the receiver telescope does not limit system operation, its performance is not expected to match that of the transmitter BEX. It consists of two plano-convex lenses with focal lengths of 150~mm (50~mm diameter) and 50~mm (25~mm diameter), providing a magnification of $M = 3$. The optical coatings support broadband operation between 650 and 1050~nm. This design prioritizes optimal performance at the qubit wavelength (795~nm), where system losses have the greatest impact on fidelity. In contrast, brightness in the classical channel can be compensated using optical amplification.

After passing through the telescope and FSM, the incoming beams are separated by the DM. The quantum beam is then filtered by a 2~nm narrowband optical bandpass filter and passed through a second telescope composed of plano-convex lenses with focal lengths of 75~mm and 25~mm. The resulting beam diameter is $\sim$5~mm. Polarization control is achieved using three motorized waveplates (quarter-wave, half-wave, quarter-wave) that apply the appropriate polarization rotations to match the transmitter’s measurement basis. A flip mount can insert a 45-degree mirror to divert the beam to a free-space–coupled polarimeter for coarse alignment using the transmitter’s auxiliary laser. Once alignment is verified, the mirror is retracted and the photonic qubits proceed to the QSA for measurement.

\subsection{Bichromatic Entanglement Source}
A key enabler of the \textit{KiQQer} experiment is the use of a bichromatic entanglement source: Qunnect’s QuSRC. Operating at room temperature, the QuSRC delivers $>10^7$ photon pairs per second with a linewidth of $\leq$1~GHz. It provides intrinsically stable polarization entanglement, strong signal-idler correlations ($g^{(2)}_{si} \gtrsim 30$), and a heralding efficiency of around 30\%~\cite{Craddock2023}. These characteristics make the QuSRC particularly well-suited for hybrid quantum networks, eliminating many limitations of traditional sources, such as low brightness, reduced fidelity, or requirements for cryogenic cooling, vacuum systems, or highly controlled laboratory environments.

The source employs a diamond-scheme spontaneous four-wave mixing (SFWM) process in warm $^{87}$Rb vapor to generate polarization-entangled photon pairs at 795~nm and 1324~nm~. Two continuous-wave lasers, a 780~nm \textit{pump} and a 1367~nm \textit{coupling} field, drive transitions from the ground state $\ket{5S_{1/2}}$ to the doubly excited state $\ket{6S_{1/2}}$. Photon-pair generation occurs via spontaneous decay through intermediate states $\ket{5P_{1/2}}$ and $\ket{5S_{1/2}}$, resulting in the emission of 1324~nm (idler) and 795~nm (signal) photons. Phase-matching ensures strong temporal and spatial correlations between the emitted pairs. When the pump and coupling fields share the same polarization, the atomic selection rules support the generation of the maximally entangled Bell state $\ket{\Phi_+} = \frac{1}{\sqrt{2}}(\ket{HH} + \ket{VV})$.

The 1324~nm photon lies within the telecom O-band, a region of low chromatic dispersion in standard single-mode fiber, making it ideal for low-loss transmission over fiber networks. In parallel, the 795~nm photon is resonant with the D1 transition of rubidium, enabling direct integration with neutral-atom–based quantum technologies such as quantum memories, clocks, and detectors. Its compatibility with atmospheric windows also makes it well suited for free-space links. This spectral separation positions the QuSRC as an effective hybrid interconnect for bridging telecom infrastructure with atomic quantum platforms, without requiring frequency conversion. The GHz-class linewidth supports efficient spectral filtering, allowing high-extinction background rejection without compromising entangled-photon quality.

The QuSRC’s brightness scales linearly with pump power, since the photon-pair creation probability depends directly on the number of pump photons participating in the nonlinear process. Higher pump powers can offset channel losses by increasing the overall pair generation rate. However, this increase introduces a trade-off: higher brightness typically reduces $g^{(2)}_{si}$ due to the rise of multi-pair events, which degrade entanglement fidelity. To balance these effects, the QuSRC was operated at a coupling laser power of $\sim$ 10mW and a pump laser power of $\sim$ 1mW -- chosen to maintain strong entanglement visibility while compensating for the high channel losses encountered from the free-space link.

\subsection{Superconducting Nanowire Single-Photon Detectors}
A fiber-coupled SNSPD system provided by Single Quantum was used to detect the 1324~nm photons emitted by the QuSRC. These detectors operate by exploiting photon-induced local superconductivity breaking within nanowires that are DC-biased close to their critical current \cite{SNSPDs2001}. This event results in a voltage pulse that is then amplified and registered as a detection event.

To enhance optical absorption, the superconducting film is deposited on top of an optical layer stack, such as a Distributed Bragg Reflector (DBR), engineered for high absorption at the target wavelength. For fiber-coupled operation, the superconducting layer is patterned into a meander-shaped nanowire covering an area slightly larger than the fiber core, maximizing coupling efficiency. The detectors used in this experiment were fabricated from NbTiN and cooled to $\sim$3~K using a Gifford–McMahon closed-cycle cryocooler. The cryostat includes fiber feedthroughs and RF cabling for both biasing and readout. Operation is controlled through an electronic interface with a user-friendly software environment.

Single Quantum’s SNSPDs are well suited for quantum communication and computing applications due to their high system detection efficiency (greater than 90\% \cite{Chang2021}), low dark count rates (below 10~cps), and extremely low timing jitter (less than 15~ps \cite{Zadeh2020}). The best-performing detector in this setup exhibited a switching current of 35.7~$\mu$A. When operated at a bias current of 33.3~$\mu$A, it achieved a 1310~nm detection efficiency of $91 \pm 3\%$, a dark count rate below 2~cps, and a timing jitter of 13~ps.

While SNSPD detection efficiency declines at high photon fluxes due to the detector's finite recovery time, this detector maintained more than 80\% efficiency up to count rates of $\sim$2.5~Mcps. This makes it an effective solution for high-rate quantum networking applications.

\subsection{Qubit Timetag Modules}
The Qubit Timetag Module (QTM) developed by Qblox enables high-resolution timestamping and event counting of photodetector signals, supporting qubit readout in optical quantum systems. Integrated into the Qblox Cluster platform, the QTM extends the system’s digital signal processing capabilities by adding time-resolved detection functions. Each module provides eight configurable channels that can operate as either inputs or outputs.

\begin{figure}[h!]
\centering
\includegraphics[width=\linewidth]{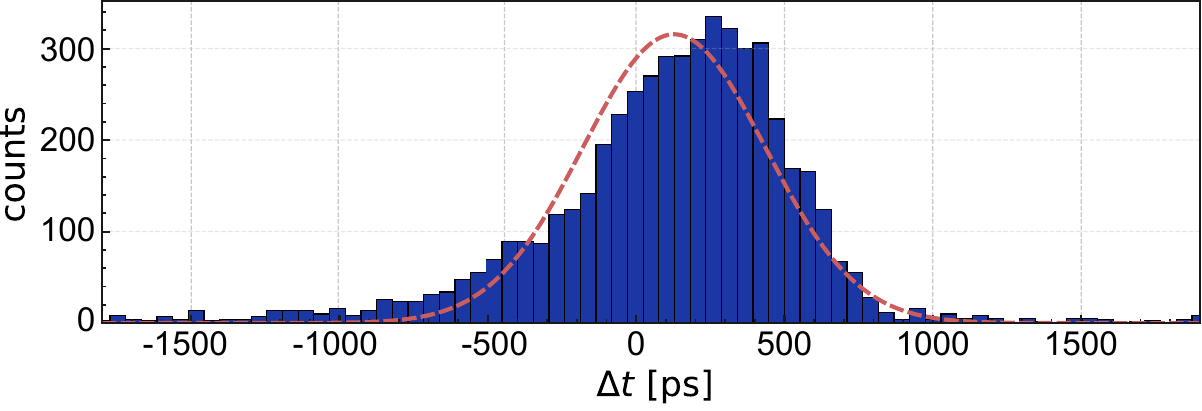}
\caption{Signal–idler correlation peak between Nodes~C and~B over a fiber link. Timing data recorded by the remotely connected QTMs was post-processed to extract coincidences.}
\label{fig:QTM_peaks}
\end{figure}

To validate QTM performance, signal–idler correlation measurements were conducted between Nodes B-C using a low-noise, low-brightness regime of the entangled photon-pair source. Under these conditions, the event rate was tuned to match the QTM’s available memory bandwidth, yielding the correlation peaks shown in Figure~\ref{fig:QTM_peaks}. The detection systems at both nodes were physically separated, and coincident events were identified by comparing locally recorded timestamps from each device. The QTM is optimized for experiments involving low photon flux and deterministic detection windows. It maintains a finite buffer of recorded events, with readout control signals and photon detections handled in a time-sequenced manner. This design enables efficient rejection of background events falling outside the expected temporal windows. Although the hardware supports multi-megahertz detection rates, the beta firmware version available at the time did not permit continuous operation at maximum throughput required for sustained use in this experiment.

\subsection{White Rabbit Time Modules}
The time synchronization modules developed by OPNT enable sub-nanosecond clock distribution between static nodes with high stability and precision. These modules support direct integration with standard C-band SFP transceivers, facilitating seamless deployment across telecom-grade infrastructure. In the \textit{KiQQer} experimental setup, wavelength-division multiplexing of the 1324~nm idler photons and the WR timing signal allows simultaneous quantum and classical communication over a single installed single-mode fiber link connecting Nodes~C and~A. This configuration provides robust, high-precision synchronization without the need for dedicated timing fibers or additional infrastructure.

\subsection{Secure Authenticated Classical Channel}
The \textit{KiQQer} project employs a Fortaegis-based solution to address the authentication requirement in QKD. Without device-level authentication, the security guarantees of QKD can be compromised, as adversaries could impersonate legitimate nodes and intercept or manipulate the key exchange. Classical authentication protocols—such as elliptic-curve cryptography (ECC)~\cite{bernstein2008binary} and RSA~\cite{rivest1978method}—rely on computational hardness assumptions that are vulnerable to quantum algorithms like Shor’s algorithm~\cite{shor1999polynomial}, which can efficiently factor large integers and compute discrete logarithms.

Fortaegis mitigates this vulnerability using a cryptographic protocol based on controlled Physical Unclonable Functions (cPUFs)~\cite{gao2020physical}. These hardware-based primitives provide unique, device-specific responses to input challenges, enabling mutual authentication, ephemeral key agreement, and automated key rotation. Prior to deployment, a bootstrap phase is used to securely exchange a set of challenge–response pairs (CRPs) among networked devices. Each CRP is composed of a challenge \( C \) and its associated response \( R \), uniquely generated by the device's cPUF due to intrinsic variations in its physical construction.

To protect CRP integrity and confidentiality, each device encrypts received CRPs of its peers using keys derived from its own cPUF responses. Decryption requires re-challenging the same cPUF under controlled conditions, enforcing hardware-bound security. This prevents CRP access in the absence of the physical device and eliminates the need for long-term keys, centralized key management, or external certificate authorities. Upon successful challenge–response exchange, a secure channel is established, mutual authentication is completed, and a fresh CRP is generated for future sessions. Because new CRPs are mathematically uncorrelated with previous ones, the approach guarantees forward secrecy and enables robust key rotation. Cryptographic material derived from the cPUF is never directly accessible to software; instead, it is confined to hardware-enforced cryptographic operations such as authentication, encryption, and key derivation using transient secrets. This architecture is based on the flowchart model presented in~\cite{vskoric2010flowchart}.

The Fortaegis engine supports multiple network configurations, including 1~Gbps, 10~Gbps, and 100~Gbps links. In the \textit{KiQQer} demonstration, devices were authenticated and connected via a 1~Gbps CAT6 copper cable. This setup established a bidirectionally encrypted and authenticated communication channel as a prerequisite for QKD session initialization.

\subsection{Quantum Time Transfer}
Picosecond-level clock offset measurements between network nodes can be realized using entangled photon pairs~\cite{3}. In the context of QTT, the ability to distribute entanglement over operational free-space optical links is essential, as global-scale timing networks require space-based links to achieve intercontinental reach~\cite{2,5,6,7}. When the entangled photon pairs are also correlated in polarization, the timing signals gain an intrinsic authentication mechanism. The impossibility of cloning a single photon’s polarization state, combined with the fundamental randomness of the temporal correlations, renders the transmission inherently resistant to spoofing and eavesdropping.

Here we show that Xairos' could implement this approach in a hybrid fiber/free-space architecture, enabling secure, high-precision clock synchronization across both terrestrial and satellite-based links~\cite{1}. This dual-channel strategy provides resilience to link interruptions and supports integration into future space-enabled timing networks.

\end{document}